\documentclass[aps,prb,twocolumn,showpacs]{revtex4}

\usepackage{graphicx}

\makeatletter
\def \frontmatter@thefootnote{\@alph\c@footnote)}%
\makeatother

\begin{document}

\title{Rigid vortices in MgB$_2$}

\author{Hao Jin}
\email{hjin@ssc.iphy.ac.cn}

\author{Hai-Hu Wen}
\email{hhwen@aphy.iphy.ac.cn}

\author{Hai-Peng Yang}

\author{Zhi-Yong Liu}

\author{Zhi-An Ren}

\author{Guang-Can Che}

\author{Zhong-Xian Zhao}

\affiliation{National Laboratory for Superconductivity, Institute of
Physics, Chinese Academy of Sciences, P.~O.~Box 603, Beijing
100080, P.~R.~China}

\date{\today}

\begin{abstract}
Magnetic relaxation of high-pressure synthesized MgB$_2$ bulks
with different thickness is investigated. It is found that the
superconducting dia-magnetic moment depends on time in a
logarithmic way; the flux-creep activation energy decreases
linearly with the current density (as expected by Kim-Anderson
model); and the activation energy increases linearly with the
thickness of sample when it is thinner than about 1 mm. These
features suggest that the vortices in the MgB$_2$ are rather
rigid, and the pinning and creep can be well described by
Kim-Anderson model.


\end{abstract}

\maketitle

The recently discovered type-II superconductor magnesium diboride (MgB$_2$)
\cite{Nagamatsu,Finnemore} exhibits noticeably different magnetic properties and flux
dynamics as compared with the high-temperature cuprate superconductors and
conventional superconductors. As a $s$-wave superconductor \cite{Kohen,Plecenik},
it shows odd behaviors in the temperature dependence of the lower critical field $H_{c1}(T)$
and the penetration depth $\lambda(T)$ \cite{Li}; the irreversibility field
$H_{\rm irr}(T)$ is remarkably lower than the upper critical field $H_{c2}(T)$ near
zero temperature; the evidently low magnetic relaxation rate depends weakly on the
temperature in the low and intermediate temperature region \cite{Wen1,Wen2}, which
makes the magnetic moment rather stable \cite{Thompson}, and it is interesting that the
relaxation rates of the high-pressure synthesized bulk and the film \cite{Wen1,Zhao1}
differ by about one order of magnitude, implying that the flux-creep activation energy
might depend on the thickness of the sample.

In view of these, the systematic magnetic relaxation
measurements of MgB$_2$ bulks in the shape of cylinder are carried
out with a superconducting quantum interference device (SQUID)
in this work, where the height (thickness) of the cylinder is
decreased at each time. It is found that the superconducting
dia-magnetic moment depends on time in a logarithmic way,
and the exponent $\mu$ in the universal equation of activation
energy \cite{Blatter,Yeshurun} is roughly $-1$ as expected by
the Kim-Anderson model for a vortex segment with a fixed length.

The high-pressure synthesized bulk sample of MgB$_2$, which is in
the shape of a cylinder with the diameter of 1.5 mm and the
original height of 3 mm, was fabricated at 950$^\circ$C for 0.5
hour under the pressure of 6 GPa. The synthesis procedures were
published elsewhere\cite{Ren}. The X-ray diffraction (XRD)
analysis shows that it is nearly in a single phase with the second
phase less than 1 wt.\%. The height of the cylinder is decreased
by sawing and polishing after each experiment, from the original
value 3 mm to 0.16 mm at last.

The measurements of magnetic relaxation are carried out by a SQUID
(Quantum Design MPMS, 5.5 T), with the temperature varying from
2 K to 32 K and in the applied magnetic field of 0.5 T , 1 T and
1.5 T, respectively. The axis of the cylinder is aligned with the
field direction during the magnetic measurements.

\begin{figure}[h!tbp]
\includegraphics[width=8cm]{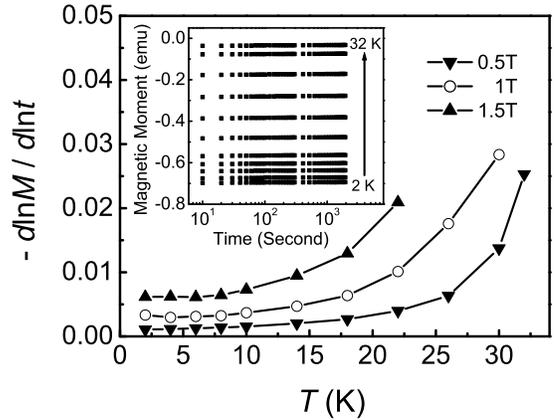}
\caption{\label{fig:Mt}
Relaxation rate of high-pressure synthesized MgB$_2$ bulk with the thickness
of 1.15 mm, the original data in the magnetic field of 0.5 Tesla are shown in
the inset, at temperatures of 2 K, 4 K, 6 K, 8 K, 10 K, 14 K, 18 K, 22 K, 26 K,
30 K and 32 K.}
\end{figure}

The data of the magnetic relaxation for the sample with thickness
of 1.15 mm and in the magnetic field of 0.5 Tesla, are plotted in
the inset of Fig.~\ref{fig:Mt}. It shows that the absolute value
of the dia-magnetic moment $M$ decreases in a logarithmic way with
time, i.~e., $M \propto \ln t$.

We start from the model of thermally activated flux motion (TAFM)
in which the dissipation is described by

\begin{equation}
E=v_0B \exp \Big(-\frac{U(j_s,T)}{k_{\rm B}T}\Big),
\end{equation}
where $j_s$ is the superconducting current density, $E$ is the electric
field due to TAFM over the activation energy $U(j_s,T)$, $v_0$ is the
average velocity of the flux motion, $B$ is the magnetic induction. For
the activation energy $U(j_s,T)$, many models have been proposed
\cite{Anderson,CP,Fisher,Zeldov}, among which the collective pinning
\cite{CP} and vortex glass \cite{Fisher} models have received
most attention in the study on the flux dynamics of cuprate superconductors.
To guarantee that $U(j_s,T)|_{j_s \rightarrow j_c} = 0$, the following
interpolation was proposed \cite{Malozemoff} and has been widely used:

\begin{equation}
U(j_s,T)=\frac{U_c(T)}{\mu}\Big[\Big(\frac{j_c(T)}{j_s}\Big)^\mu-1\Big]
\label{eq:ujT}
\end{equation}
where $U_c$ is the activation energy of vortices. Although Eq.~(\ref{eq:ujT})
was proposed based on the flux dynamics in cuprate superconductors, it can
be used as a general equation which describes the flux dynamics under specific
situation by choosing different $\mu$ values. For example the linear $U(j)$
dependence requested by the Kim-Anderson model \cite{Anderson}
to describe the creep of a vortex segment with fixed length can be
obtained by choosing $\mu=-1$. From the above two equations, one can
derive the superconducting current density,

\begin{equation}
j(t)\simeq j_c \Big[1+\frac{\mu k_{\rm B}T}{U_c}
\ln \Big(1+\frac{t}{t_0}\Big)\Big]^{-\frac{1}{\mu}}.\label{eq:jt}
\end{equation}

When choosing $\mu=-1$, one can easily see that $j$ depends on
time in a logarithmic way which is consistent with the data shown
in the inset of Fig.~\ref{fig:Mt}. This feature may suggest that the
flux dynamics can be described by the Kim-Anderson model rather
than the collective pinning model as suggested recently
\cite{Qin}. The temperature dependence of the relaxation
rate $S=-d \ln M/d \ln t$ is then derived, as shown in
Fig.~\ref{fig:Mt}. The extrapolation of relaxation rate does not
approach zero as $T \rightarrow 0$, which has been interpreted
by either quantum tunneling of vortices or the two-gap feature of
MgB$_2$  \cite{Wen1,Wen2}.

\begin{figure}[h!tbp]
\includegraphics[width=8cm]{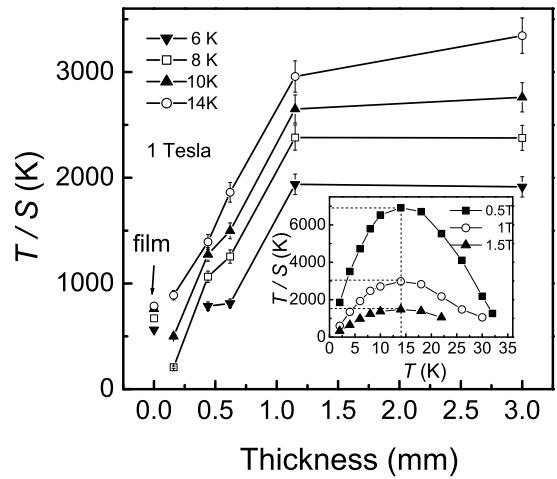}
\caption{\label{fig:T/S}Thickness dependence of the activation energy $U_c(T)$
(approximated by $T/S$) at 1 Tesla. $U_c$ is proportional to the thickness of sample,
and saturates at about 1 mm. The leftmost isolated dots are from the data of the film
\cite{Wen1,Wen2} with the thickness of 400 nm. The inset shows $T/S$ versus $T$
with the same data.}
\end{figure}

In order to get the activation energy, we use the relation \cite{Wen3}

\begin{equation}
T/S=U_c/k_{\rm B}+\mu T \ln(1+t/t_0),\label{eq:Uc}
\end{equation}
where $\ln(1+t/t_0)$ can be treated as a constant within a fixed
time window. The inset of Fig.~\ref{fig:T/S} shows a big hump
in each curve of $T/S$ versus $T$. It is shown in Fig.~\ref{fig:Mt} that
the relaxation rate $S$ keeps flat up to a temperature of about 15 K.
Therefore the rising part of the big hump in the low temperature
region is induced by the weak temperature dependence of $S$. When
$T$ is above 15 K, $T/S$ decreases with temperature which may be
induced by the decreasing of $U_c(T)$ or by a negative $\mu$
value. One can have an approximate value for $U_c$(15 K), which is
about 7000 K for 0.5 T. This value is about 10 times greater than
that in the high-temperature cuprate superconductors. We will show below
that this larger activation energy is induced by the strong pinning to a very
long rigid vortex.

\begin{figure}[h!tbp]
\includegraphics[width=8cm]{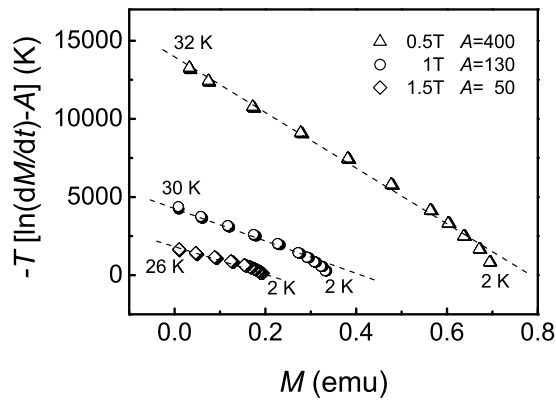}
\caption{\label{fig:Maley} The dependence of the flux-creep activation energy on the
persistent current density determined by Maley's method. The relevant temperatures
(from the right to the left) are 2K, 4K, 6 K, 8 K, 10 K, 14 K, 18 K, 22 K, 26 K, 30 K,
and 32 K.}
\end{figure}

One important factor characterizing the vortex dynamics is the
$U(j)$ dependence. For collective pinning, $U(j)$ is non-linear and
the $\mu$ value in Eq.~(\ref{eq:ujT}) is positive. However, as forementioned, for a
vortex segment with fixed length, Kim-Anderson model proposes a
linear $U(j)$ relation. In our present work, Maley's method \cite{Yeshurun,Maley}
is used to determine the dependence of the activation energy on the current density.
The results are shown in Fig.~\ref{fig:Maley}. The linear $U(j)$ relation suggests
an Kim-Anderson type effective pinning barrier $U(j) =U_c(1-j/j_c)$ with $\mu=-1$.
At low temperatures, the curves deviate from linearity, which indicates that other
mechanism may set in, leading to the rather stable $S$ value in low temperature
region. It is interesting to note that in high-temperature cuprate
superconductors, such as YBa$_2$Cu$_3$O$_7$, $U(j)$ diverges in the small
current limit which is resulted by the elastic property of the vortices.

Based on the assumption $U_c(T) \approx T/S$, we calculate the
values of $U_c$. Fig.~\ref{fig:T/S} shows that the
values of $U_c$ is proportional to the thickness of sample
and saturates at about 1 mm. It also depends on the disorder
strength which is different between bulks and films \cite{Zhao2}.
In the scenario of weak collective pinning where the pinning
barrier is caused by the weak point disorder, vortices will
break up into segments \cite{Blatter} due to the competition
between the elastic and pinning energy, the longitudinal size of
which is the collective pinning length $L_c \simeq(\varepsilon_0^2\xi^2/ \gamma)^{1/3}$,
with $\varepsilon_0$ the basic energy scale, $\xi$ the coherence
length, and $\gamma$ a parameter of disorder strength. Thus the
linear dependence of $U_c$ on the sample thickness suggests
that the flux line in MgB$_2$ bulks will behave as an individual
segment if its length is below $L_c \simeq 1$ mm. Combining with
the significant large value of $U_c$ ($\simeq 10^3$ K), we conclude
that the vortices are remarkably rigid over the whole sample.

For a long and rigid vortex, the intrinsic pinning energy $U_c$
is large, therefore the thermal activation barrier is high, vortices
may move to the next pinning site by quantum tunneling.
This naturally interprets the rather stable $S$ in the temperature region
up to 15 K ($\simeq 0.4 T_c$) and also the large gap between
$H_{c2}(0)$ and $H_{\rm irr}(0)$. However, the big gap between
$H_{c2}(0)$ and $H_{\rm irr}(0)$ may be explained by an alternative
model based on two-gap feature $\Delta_1(0)=1.8$ meV and
$\Delta_2(0)=6.8$ meV \cite{Choi}, that the upper critical field $H_{c2}(T)$
is determined by $\Delta_2(T)$, while the irreversibility field
$H_{\rm irr}(T)$ may include the contribution of both gaps.
The thickness dependence of $U_c$ even holds below 15 K, implying that some
further research are needed to clarify whether the quantum tunneling or the
two-gap feature dominates.

In conclusion, the magnetic relaxation of high-pressure synthesized MgB$_2$
cylinder is measured by SQUID in a series of experiments, with the height of
sample decreased each time. The linear $M$-$\ln t$, $U$-$j$ relations and the thickness
dependence of the activation energy are found. These features imply a Kim-Anderson
type flux creep and the rather rigid vortices with the collective pinning length $L_c$ about
1 mm in bulk MgB$_2$, in contrast to the cuprate superconductors and the previous
results of collective-pinning. The thickness dependence of $U_c$ even at low temperatures
also suggest the quantum tunneling to be still as an open question.

This work is financially supported by the National Natural Science Foundation
of China (Grant No.~NSFC 19825111) and the Ministry of Science and Technology
of China (Grant No.~NKBRSF-G1999064602).

\end{document}